\documentclass{elsart5p}

\usepackage{latexsym}
\usepackage{graphicx}
\usepackage{amsmath}
\usepackage{amsfonts}   
\usepackage{amssymb}    

\def\mt{{\ifmmode M^{eff}_T\else $M^{eff}_T$\fi}}

\def\e{\epsilon}

\def\ra{\rangle}

\def\e3{$\epsilon_3$}

\def\ch2{$\chi2$}

\def\co#1{{\ifmmode{\cal O}_{#1}\else${\cal O}_{#1}$\fi}}

\newdimen\unit
\def\point#1 #2 #3{\vbox to0pt{\kern-#2\unit
  \hbox{\kern#1\unit#3}\vss}
 \nointerlineskip}
\newcommand{\be}{\begin{equation}}
\newcommand{\ee}{\end{equation}}
\newcommand{\bea}{\begin{eqnarray}}
\newcommand{\eea}{\end{eqnarray}}

\newcommand{\mev}{\mbox{ MeV }}
\newcommand{\gev}{\mbox{ GeV}}
\newcommand{\tev}{\mbox{ TeV}}
\newcommand{\GeV}{\gev}

\newcommand{\TeV}{\tev}
\newcommand{\cl}{\text{CL}}

\newcommand{\dasusy}{\delta a_{\mu}^{\text{SUSY}}}
\newcommand{\alphaemmz}{\alpha_{\text{em}}(M_Z)^{\overline{MS}}}
\newcommand{\alphas}{\alpha_s(M_Z)^{\overline{MS}}}

\newcommand{\data}{d}

\newcount\hour
\newcount\minute
\newtoks\amorpm
\hour=\time\divide\hour by60 \minute=\time{\multiply\hour by60
\global\advance\minute by- \hour}
\edef\standardtime{{\ifnum\hour<12 \global\amorpm={am}%
    \else\global\amorpm={pm}\advance\hour by-12 \fi
    \ifnum\hour=0 \hour=12 \fi
    \number\hour:\ifnum\minute<100\fi\number\minute\the\amorpm}}
\edef\militarytime{\number\hour:\ifnum\minute<100\fi\number\minute}
\def\bold#1{\setbox0=\hbox{$#1$}%
     \kern-.025em\copy0\kern-\wd0
     \kern.05em\copy0\kern-\wd0
     \kern-.025em\raise.0433em\box0 }

\newcommand{\newc}{\newcommand}
\newc\eg{{\it {e.g.~}}}
\newc\ie{{\it {i.e.}}}
\newc\etc{{\it {etc}}}
\newcommand\lsim{\mathrel{\rlap{\lower4pt\hbox{\hskip1pt$\sim$}}
    \raise1pt\hbox{$<$}}}
\newcommand\gsim{\mathrel{\rlap{\lower4pt\hbox{\hskip1pt$\sim$}}
    \raise1pt\hbox{$>$}}}
\newc{\mhalf}{m_{1/2}}      \newc{\mzero}{m_0}
\newc{\tanb}{\tan\beta}
\newc{\azero}{A_0}
\newc{\at}{A_t} \newc{\ab}{A_b} \newc{\atau}{A_\tau}
\newc{\bmu}{B\mu}           \newc{\sgn}{{\rm sgn}}
\newc{\mone}{M_1}           \newc{\mtwo}{M_2}
\newc{\charone}{\chi_1^\pm} \newc{\mcharone}{m_{\chi_1^\pm}}
\newc{\hl}{h}               \newc{\mhl}{m_{\hl}}
\newc{\hh}{H}               \newc{\mhh}{m_{\hh}}
\newc{\ha}{A}               \newc{\mha}{m_{\ha}}
\newc{\hc}{H^{\pm}}         \newc{\mhc}{m_{\hc}}
\newc{\qzero}{Q_0}          \newc{\qstop}{Q_{\widetilde t}}
\newc{\amu}{a_{\mu}}        \newc{\amususy}{a_{\mu}^{\rm SUSY}}
\newc{\amuexpt}{a_{\mu}^{\rm expt}}        \newc{\amusm}{a_{\mu}^{\rm SM}}
\newc{\deltaamususy}{\Delta a_{\mu}^{\rm SUSY}}
\newc\gmtwo{(g-2)_{\mu}} \newc\deltaamu{\Delta a_{\mu}}
\newc{\msbar}{\overline {\rm MS}} \newc{\drbar}{\overline {\rm DR}}
\newc{\yt}{h_t} \newc{\yb}{h_b} \newc{\ytau}{h_{\tau}}
\newc{\mtpole}{M_t}
\newc{\mtaupole}{m_{\tau}^{\rm pole}}
\newc{\mtmtsmmsbar}{m_t(m_t)^{\msbar}_{{\rm SM}}}
\newc{\mtmtsmdrbar}{m_t(m_t)^{\drbar}_{{\rm SM}}}
\newc{\mtmtmssmdrbar}{m_t(m_t)^{\drbar}_{{\rm SUSY}}}
\newc{\mbmbsmmsbar}{m_b(m_b)^{\msbar}_{{\rm SM}}}
\newc{\mbmzsmmsbar}{m_b(\mz)^{\msbar}_{{\rm SM}}}
\newc{\mbmzsmdrbar}{m_b(\mz)^{\drbar}_{{\rm SM}}}
\newc{\mbmzmssmdrbar}{m_b(\mz)^{\drbar}_{{\rm SUSY}}}
\newc{\mtaumzsmmsbar}{m_{\tau}(\mz)^{\msbar}_{{\rm SM}}}
\newc{\mtaumzsmdrbar}{m_{\tau}(\mz)^{\drbar}_{{\rm SM}}}
\newc{\mtaumzmssmdrbar}{m_{\tau}(\mz)^{\drbar}_{{\rm SUSY}}}
\newc{\mgut}{M_{\rm GUT}}
\newc{\mplanck}{M_{\rm P}}      \newc{\mpl}{M_{\rm Pl}}
\newc{\msusy}{M_{\rm SUSY}}      \newc{\ms}{M_{\rm S}}
\newc{\jxf}{J({\xf})}
\newc{\jxfexact}{J_{\rm exact}({\xf})}  \newc{\jxfexp}{J_{\rm exp}({\xf})}
\newc{\VEV}[1]{\langle #1 \rangle}
\newc{\xf}{x_f}
\newc\vrel{v_{\rm rel}}
\newcommand\mchi{m_{\chi}}              
\newc\sell{{\widetilde e}_L}      \newc\msell{m_{\sell}}
\newc\selr{{\widetilde e}_R}      \newc\mselr{m_{\selr}}
\newc\snue{{\widetilde \nu}_e}      \newc\msnue{m_{\snue}}
\newc\snutau{{\widetilde \nu}_\tau}      \newc\msnutau{m_{\snutau}}
\newc\supl{{\widetilde u}_L}      \newc\msupl{m_{\supl}}
\newc\supr{{\widetilde u}_R}      \newc\msupr{m_{\supr}}
\newc\sdl{{\widetilde d}_L}      \newc\msdl{m_{\sdl}}
\newc\sdr{{\widetilde d}_R}      \newc\msdr{m_{\sdr}}

\newc\hpm{H^\pm} \newc\hp{H^+} \newc\hm{H^-}
\newc\sfermion{\tilde f}  \newc\msfermion{m_{\sfermion}}
\newc\kmeter{{\rm km}}
\newc\second{{\rm sec}}
\newc{\gstar}{g_\ast}           \newc{\gsstar}{g_{s\ast}}
\newc{\geff}{g_{\rm eff}}
\newcommand\mz{m_{Z}}

\newc{\sthw}{\sin\theta_W}              \newc{\cthw}{\cos\theta_W}
\newc{\bino}{\widetilde B}              \newc{\wino}{\widetilde W_30}
\newc{\higgsinob}{{\widetilde H}^0_b}   \newc{\higgsinot}{{\widetilde H}^0_t}
\newc{\abund}{\Omega h^2}
\newc{\abundchi}{\Omega_\chi h^2}
\newc{\abundcdm}{\Omega_{\text{CDM}} h^2}
\newc{\omegam}{\Omega_{M}}       \newc{\abundm}{\Omega_{M} h2}
\newc{\omegab}{\Omega_{b}}       \newc{\abundb}{\Omega_{b} h2}
\newc{\omegacdm}{\Omega_{CDM}}
\newc{\omegatot}{\Omega_{TOT}}
\newc{\rhocrit}{\rho_{crit}}
\newc{\rhochi}{\rho_{\chi}}
\newcommand\pb{\,\mbox{pb}}

\newc\br{\mbox{BR}}

\newc{\beq}{\begin{equation}}
\newc{\eeq}{\end{equation}}

\newc\stoponetwo{{\widetilde t}_{1,2}}
\newc\sbotonetwo{{\widetilde b}_{1,2}}
\newc\stauonetwo{{\widetilde \tau}_{1,2}}

\newc\bsgamma{b\ra s \gamma }
\newc\brbsgamma{\br( B\rightarrow X_s \gamma )}
\newc{\sigsip}{\sigma^{SI}_{p}} \newc{\sigsin}{\sigma^{SI}_{n}}
\newc{\sigsiN}{\sigma^{SI}_{N}}
\newc{\sigsdp}{\sigma^{SD}_{p}} \newc{\sigsdn}{\sigma^{SD}_{n}}
\newc{\sigsiA}{\sigma^{SI}_{A}}

\newc\xilim{\xi_{\rm lim}} 
\newc\tlim{t_{\rm lim}} 
\newc\zetalim{\zeta_{\rm lim}} 

\newc\zetah{\zeta_h}
\newc{\relprobone}[1]{p({#1} \vert d)}
\newc{\relprobtwo}[2]{p({#1},{#2} \vert d)}

\long\def\begincomment#1\endcomment{%
        \begingroup\sf\baselineskip12pt#1\endgroup}

\newc\AJ[3]
                {{\em Astrophys. J.} {\bf #1} (#2) #3}
\newc\AP[3]
                {{\em Ann. Phys.} {\bf #1} (#2) #3}
\newc\APJ[3]
                {{\em Astropart. J.} {\bf #1} (#2) #3}
\newc\APP[3]
                {{\em Astropart. Phys.} {\bf #1} (#2) #3}
\newc\APS[3]
                {{\em Astrophys. J. Suppl.} {\bf #1} (#2) #3}
\newc\ARNPS[3]
                {{\em Ann. Rev. Nucl. Part. Sci.} {\bf C#1} (#2) #3}
\newc\CPC[3]
                {{\em Comput. Phys. Commun.} {\bf C#1} (#2) #3}
\newc\EPJ[3]
                {{\em Euro. Phys. Journ.} {\bf C#1} (#2) #3}
\newc\JCAP[3]
                {{\em JCAP} {\bf #1} (#2) #3}
\newc\JHEP[3]
                {{\em JHEP} {\bf #1} (#2) #3}
\newc\IJMP[3]
                {{\em Int. J. Mod. Phys.} {\bf A#1} (#2) #3}
\newc\MNRAS[3]
                 {{\em Mon. Not. Roy. Astron. Soc.} {\bf #1} (#2) #3}
\newc\MPL[3]
                {{\em Mod. Phys. Lett.} {\bf A#1} (#2) #3}
\newc\NCA[3]
                {{\em Nuovo Cimento} {\bf #1} (#2) #3}
\newc\NIM[3]
                {{\em Nucl. Instrum. Methods} {\bf #1} (#2) #3}
\newc\NIMA[3]
                {{\em Nucl. Instrum. Methods} {\bf A#1} (#2) #3}
\newc\NAT[3]
                {{\em Nature} {\bf #1} (#2) #3}
\newc\NPB[3]
                {{\em Nucl. Phys.} {\bf B#1} (#2) #3}
\newc\PL[3]
                {{\em Phys. Lett.} {\bf B#1} (#2) #3}
\newc\PLB[3]
                {{\em Phys. Lett.} {\bf B#1} (#2) #3}
\newc\PR[3]
                {{\em Phys. Rep.} {\bf #1} (#2) #3}
\newc\PRL[3]
                {{\em Phys. Rev. Lett.} {\bf B#1} (#2) #3}
\newc\PRD[3]
                {{\em Phys. Rev.} {\bf D#1} (#2) #3}
\newc\PTP[3]
                {{\em Prog. Theor. Phys.} {\bf #1} (#2) #3}
\newc\RMP[3]
                {{\em Rev. Mod. Phys.} {\bf #1} (#2) #3 }
\newc\RPP[3]
                {{\em Rept. Prog. Phys.} {\bf #1} (#2) #3 }
\newc\SC[3]
                {{\em Science} {\bf #1} (#2) #3 }
\newc\ZPC[3]
                {{\em Z. Phys.} {\bf C#1} (#2) #3}
\newc\Err[3]
           {{\em Erratum-ibid.} {\bf #1} (#2) #3 }

\newcommand{\squishlist}{
   \begin{list}{$\bullet$}
    { \setlength{\itemsep}{0pt}      \setlength{\parsep}{3pt}
      \setlength{\topsep}{3pt}       \setlength{\partopsep}{0pt}
      \setlength{\leftmargin}{1.em} \setlength{\labelwidth}{1em}
      \setlength{\labelsep}{0.5em} } }
\newcommand{\squishend}{
    \end{list}  }
        
\newcommand{\nuis}{\psi} 
\newcommand{\params}{\theta} 
\newcommand{\basis}{\eta}

\newcommand{\sineff}[1]{\protect{\sin^2\theta_{\text{eff}}^{#1}}}

\begin{document}

\begin{frontmatter}



\title{Prospects for direct dark matter detection in the Constrained MSSM}


\author[rt]{Roberto Trotta}
\address[rt]{Astrophysics Department, Oxford University \\
        Denys Wilkinson Building,  Keble Road, Oxford OX1 3RH, United Kingdom}
\author[rra]{Roberto Ruiz de Austri}
\address[rra]{
        Departamento de F\'{\i}sica Te\'{o}rica C-XI
        and Instituto de F\'{\i}sica Te\'{o}rica C-XVI,\\
        Universidad Aut\'{o}noma de Madrid, Cantoblanco,
        28049 Madrid, Spain}
\author[lr]{Leszek Roszkowski}
\address[lr]{Department of Physics and Astronomy, University of Sheffield,\\
        Sheffield S3 7RH, England}

\begin{abstract}
We outline the WIMP dark matter parameter space in the Constrained
MSSM by performing a comprehensive statistical analysis that
compares with experimental data predicted superpartner masses and
other collider observables as well as a cold dark matter
abundance. We include uncertainties arising from theoretical
approximations as well as from residual experimental errors on
relevant Standard Model parameters.

We present high--probability regions for neutralino dark matter
direct detection cross section, and we find that $ 10^{-10}\pb
\lsim \sigsip \lsim 10^{-8}\pb$ for direct WIMP detection (with
details slightly dependent on the assumptions made). We highlight
a complementarity between LHC and WIMP dark matter searches in
exploring the CMSSM parameter space. We conclude that most of the
$95\%$ probability region for the cross section will be explored
by future one--tonne detectors, that will therefore cover most of
the currently favoured region of parameter space.
\end{abstract}

\begin{keyword}

\PACS
\end{keyword}
\end{frontmatter}

\section{Introduction}\label{sec:intro}

Two of the most challenging questions facing particle physics
today are the instability of the Higgs mass against radiative
corrections (known as the ``fine--tuning problem'') and the nature
of dark matter. Unlike the Standard Model (SM), weak scale softly
broken supersymmetry (SUSY) provides solutions to both of them.
Firstly, the fine--tuning problem is addressed via the
cancellation of quadratic divergences in the radiative corrections
to the Higgs mass.  Secondly, assuming $R$--parity, the lightest
supersymmetric particle (LSP) is a leading weakly interactive
massive particle (WIMP) candidate for cold dark matter (CDM).
Despite these and other attractive features, without a reference
to grand unified theories (GUTs), low energy SUSY models suffer
from the lack of predictivity due to a large number of free
parameters (\eg, over 120 in the Minimal Supersymmetric Standard
Model (MSSM)), most of which arise from the SUSY breaking sector.

The MSSM with one particularly popular choice of universal
boundary conditions at the grand unification scale is called the
Constrained Minimal Supersymmetric Standard Model
(CMSSM)~\cite{kkrw94}. The CMSSM is defined in terms of five free
parameters: common scalar ($\mzero$), gaugino ($\mhalf$) and
tri--linear ($\azero$) mass parameters (all specified at the GUT
scale) plus the ratio of Higgs vacuum expectation values $\tanb$
and $\text{sign}(\mu)$, where $\mu$ is the Higgs/higgsino mass
parameter whose square is computed from the conditions of
radiative electroweak symmetry breaking (EWSB). The economy of
parameters in this scheme makes it a useful tool for exploring
SUSY phenomenology.

Many studies have explored the phenomenology of the CMSSM or other
SUSY models, mostly by evaluating the goodness--of--fit of points
scanned using fixed grids in parameter space, see
\eg~\cite{grid1,grid2,grid3,grid4,grid5,grid6}. However, this
approach has several severe limitations.  Firstly, the number of
points required scales as $k^N$, where $N$ is the number of the
model's parameters and $k$ the number of points for each of them.
Therefore this approach becomes highly inefficient for exploring
with sufficient resolution  parameter spaces of even modest
dimensionality, say $N>3$. Secondly, narrow ``wedges'' and similar
features of parameter space can easily be missed by not setting a
fine enough resolution (which, on the other hand, may be
completely unnecessary outside such special regions). Thirdly,
extra sources of uncertainties (\eg, those due to the lack of
precise knowledge of SM parameter values) and relevant external
information (\eg, about the parameter range) are difficult to
accommodate in this scheme.

In this work we report results from a Bayesian exploration of the
CMSSM parameter space, obtained through the use of Markov Chain
Monte Carlo methods. In particular we focus on the prospects for
direct neutralino dark matter detection with the next generation
of dark matter searches. We refer the reader to \cite{rrt1} for
full details. The Bayesian approach has several technical and
statistical advantages over the more traditional fixed--grid scan
technique, the most important being perhaps the ability to
incorporate all relevant sources of uncertainties, \eg the
residual uncertainty in the value of SM parameters. This means
that the inferred high probability regions in terms of \eg
neutralino mass and scattering cross section take fully into
account all sources of uncertainty relevant to the problem. For
other works applying a similar approach to the CMSSM,
see~\cite{bg04} (with some relevant differences) and more recently
\cite{al05,allanach06}.

\section{Parameter space, priors and data used}

We restrict our analysis to the case $\text{sign}(\mu) = + 1$, as
motivated by the fact that the observed anomalous magnetic moment
of the muon is positive, and since the sign of the SUSY
contribution to it is the same as the sign of $\mu$. We thus
consider the 8 dimensional parameter space $(\params, \nuis)$,
where $\params$ is a vector of CMSSM parameters,
 \be \label{eq:params}
 \params= (\mzero, \mhalf, \azero, \tanb)
 \ee
while $\nuis$ is a vector of relevant SM parameters,
 \be
 \nuis =  (\mtpole,
m_b(m_b)^{\overline{MS}}, \alphaemmz, \alphas ),
 \ee
where $\mtpole$ is the pole top quark mass,
$m_b(m_b)^{\overline{MS}}$ is the bottom quark mass at $m_b$,
while $\alphaemmz$ and $\alphas$ are the electromagnetic and the
strong coupling constants at the $Z$ pole mass $M_Z$, the last
three evaluated in the $\overline{MS}$ scheme. We are not
interested in constraining the value of the SM parameters, but
rather in including the effect of the uncertainty in their
experimental determination on the mass spectra, cosmological dark
matter abundance and other observable quantities. This in turns
has a non--negligible impact when making inferences on the high
probability regions for the CMSSM parameters, $\params$, or for
any other derived quantity. At the end of the analysis, the set of
so called ``nuisance parameters'' $\nuis$ is integrated out from
our probability distribution function (pdf). It turns out that
including the nuisance parameters in our analysis as free
parameters (rather than fixing them to the central experimental
value as it is done in most grid scans) has an important impact in
widening the constraints on the CMSSM parameters, an effect that
should not be ignored when carrying out proper statistical
inference on $\params$ or on other quantities of interest such as
the dark matter scattering cross section.

Bayesian statistics makes use of Bayes theorem,
 \be
\label{eq:bayes}
 p(\eta | \data) = \frac{p(\data |
\eta, f(\eta)) \pi(\eta)}{p(\data)}. \ee
where we have introduced $\eta = (\params, \nuis)$, to compute the
posterior probability distribution $p(\eta | \data)$. On the rhs
of Eq.~\eqref{eq:bayes}, the quantity $p(\data | \eta, f(\eta))$,
is called the {\em likelihood} and it supplies the information
provided by the data, by comparing the base parameters $\eta$ or
any derived function $f(\eta)$ to the data $\data$. The quantity
$\pi(\basis)$ denotes a {\em prior probability density function}
(hereafter called simply {\em a prior}) which encodes our state of
knowledge about the values of the parameters before we see the
data. We take the prior to be flat (\ie, constant) in the
variables $\eta$, and we further need to specify the range along
each direction. If the constraining power of the likelihood is
strong enough to override the choice of the prior, than the latter
does not matter in the final inference based on the posterior pdf.
We found this to be the case for all of the parameters but $m_0$
and sparticles masses that mostly depend on it, such as sleptons
and squarks. For those quantities, choosing a prior $50\GeV \leq
m_0 \leq 2 \TeV$ cuts away a large region of parameter space that
is not disfavoured by data, the so called ``focus point region''.
If the prior range is enlarged to $50\GeV \leq m_0 \leq 4 \TeV$,
then a considerable part of the focus point region is included in
the analysis, even though it is still not possible to constrain
the value of $m_0$ from above. With presently available data, any
upper limit on $m_0$ is purely a consequence of the prior adopted.
In the following we present results for the extended range with a
prior region up to $4 \TeV$ for $m_0, \mhalf$, and in the range
$|\azero| \leq 7 \TeV$, $2 \leq \tanb \leq 62$. The prior range on
the nuisance parameters does not influence the final results,
since the SM parameters are rather tightly constrained by the data
(summarized in the top section of Table~\ref{tab:meas}).

For our analysis, from the CMSSM and SM parameters $\eta$ we
compute a series of derived observable quantities $f(\eta)$: the
$W$ gauge boson mass, the effective leptonic weak mixing angle
$\sineff{}$, the anomalous magnetic moment of the muon, $a_\mu
\equiv (g-2)_\mu $, the branching ratios $BR(\bar{B} \rightarrow
X_s \gamma)$ and $BR(B_s \rightarrow \mu^+ \mu^-)$, the
cosmological neutralino relic abundance $\abundcdm$ (by solving
the Boltzmann equation numerically as in~\cite{darksusy}), the
light Higgs mass and the superpartner masses (computed with the
package SOFTSUSY~v1.9~\cite{softsusy}). For all of those
quantities, relevant measurements (summarized in the bottom
section of Table~\ref{tab:meas}) or experimental limits are
included via the likelihood and used to constrain high posterior
probability regions of the model. We assume that all of the cold
dark matter is in the form of neutralinos. Furthermore, the
likelihood is modified in such a way that it includes estimated
theoretical uncertainties in the mapping from CMSSM  and SM
parameters to derived quantities, another major advantage of
employing a Bayesian approach (see~\cite{rrt1} for details).
\begin{table}
\centering
\begin{tabular}{|l | l l l|}
\hline
Nuisance parameter  &   Mean value  & \multicolumn{2}{c|}{Uncertainty}\\
 &   $\mu$      & experimental & theoretical   \\ \hline
$\mtpole$           &  172.7 GeV    & 2.9 GeV & N/A\\
$m_b (m_b)^{\overline{MS}}$ &4.24 GeV  & 0.11 GeV & N/A  \\
$\alphas$       &   0.1186   & 0.002 & N/A\\
$1/\alphaemmz$  & 127.958 & 0.048 & N/A \\ \hline Derived
observable & & & \\\hline
 $M_W$          &  $80.425 \gev$     & $34\mev$ & $13\mev$  \\
 $\sineff{}$    &  $0.23150$      & $16\times10^{-5}$
                &$25\times10^{-5}$   \\
$\dasusy \times 10^{10}$       &  $25.2 $ & 9.2 &  $1$  \\
 $\text{BR}(\bar{B} \rightarrow X_s \gamma) \times 10^{4}$ &
 $3.39$ & $0.30$ & $0.30$  \\
 $\abundchi$ &  0.119 & 0.009 & $0.1\,\abundchi$ \\\hline
\end{tabular}
\caption{Top section: experimental mean $\mu$ and standard
deviation for the nuisance parameters used in the analysis. Bottom
section: as above, but for derived observable quantities,
including a theoretical uncertainty describing the imprecise
mapping of CMSSM and SM parameters onto observable quantities.
Notice that the value for the cosmological neutralino abundace
$\abundchi$ stems from WMAP 1st year data combined with other
cosmological observations. \label{tab:meas}}
\end{table}
We then compute a spin--independent dark matter WIMP elastic
scattering cross section on a free proton, $\sigsip$, including
full supersymmetric contributions which have been derived by
several groups~\cite{dn93scatt:ref,jkg96,bb98,efo00,knrr1}, but we
do not include current constraints in the likelihood, in view of
the uncertainties in the structure of the Galactic halo (\eg,
existence of clumps of dark matter and therefore the value of the
local halo mass density) as well as in the values of some hadronic
matrix elements entering the computation of $\sigsip$. Our
Bayesian approach allows to easily compute the posterior pdf for
the cross section or any other derived variable. As we now
discuss, our result for high--probability regions is highly
suggestive of a possible detection by the next generation of
direct dark matter searches.

\section{Results for neutralino direct detection}

\begin{figure}[tb]
\centering
\includegraphics[width=\linewidth]{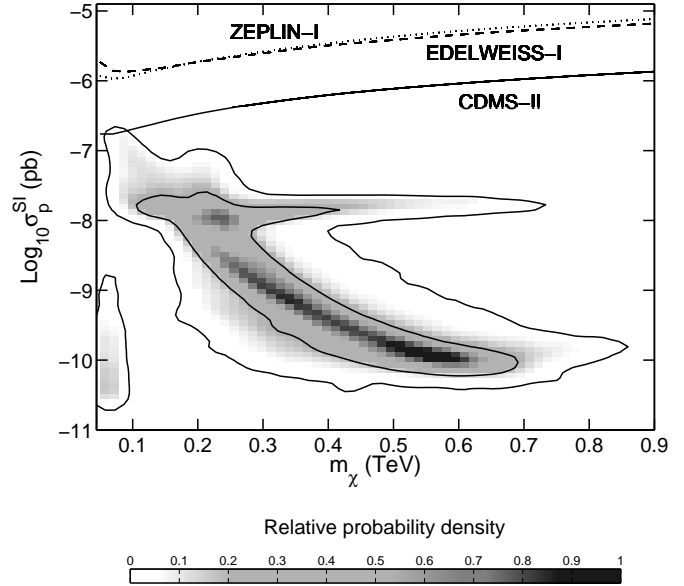}
\caption{The 2--dimensional probability density in the neutralino
mass and spin--independent cross section plane in the CMSSM (all
other parameters marginalized) with the contours containing 68\%
and 95\% probability also marked.  Current 90\% experimental upper
limits are also shown. A large fraction of the high--probability
region lies just below current constraints and it will be probed
by the next generation of dark matter searches, starting from the
focus point region (horizontal region at $\sigsip \sim 10^{-8}$).
} \label{fig:dd}
\end{figure}

In Fig.~\ref{fig:dd} we present the 2--dimensional posterior pdf
for $\sigsip$ and $\mchi$, with all other parameters marginalized
over. For comparison, we also show current
CDMS--II~\cite{cdms04limit},
Edelweiss--I~\cite{edelweiss-one-final} and UKDMC
ZEPLIN--I~\cite{zeplin-one-final} 90\%~\cl\ upper limits, but we
stress that this constraint has not been used in our analysis.

In Fig.~\ref{fig:dd} the biggest, banana--shaped region of high
probability ($68\%$ regions delimited by the internal solid, blue
curve) shows a well--defined anticorrelation between $\sigsip$ and
$\mchi$. It results from two allowed regions in the CMSSM
parameter space: the bulk and stau coannihilation region and from
the $\ha$--resonance at large $\tanb$. This region covers roughly
the range $10^{-10} \lsim \sigsip\lsim10^{-8}\pb$ and
$200\lsim\mchi\lsim700\gev$. In both cases the dominant
contribution to $\sigsip$ comes from a heavy Higgs exchange. At
small $\mchi\lsim100\gev$ we notice a small vertical band of
fairly low probability density ($\lsim0.2$) at small $\sigsip$, a
region where the light Higgs resonance contribute to reducing
$\abundchi$ at small $\mhalf$ to meet the WMAP measurement. It is
interesting to notice that our Monte Carlo procedure is able to
resolve such small features of parameter space, that are almost
invariably missed by usual fixed--grid scans. This region would
also disappear with a fair improvement in the lower bound on
$\mhl$. Finally, we can see a well pronounced region of high
probability at fairly constant $\sigsip\sim 1.6\times10^{-8}\pb$
for $\mchi\lsim420\gev$ which at low $\mchi$ partly overlaps with
the previous region. At $95\%$ this region extends up to
$\mchi\lsim720\gev$ for fairly constant $\sigsip$. This ``high''
$\sigsip$ band results from the focus point region, basically
independently of $\tanb$. This result has to be interpreted
carefully, since there are large uncertainties associated with FP
region, in particular with its location in the $(\mhalf,\mzero)$
plane mentioned earlier. Despite those outstanding questions, we
believe that it is safe to expect that the FP will be the first to
be probed by dark matter search experiments.

\begin{figure}[tb]
\centering
\includegraphics[width=\linewidth]{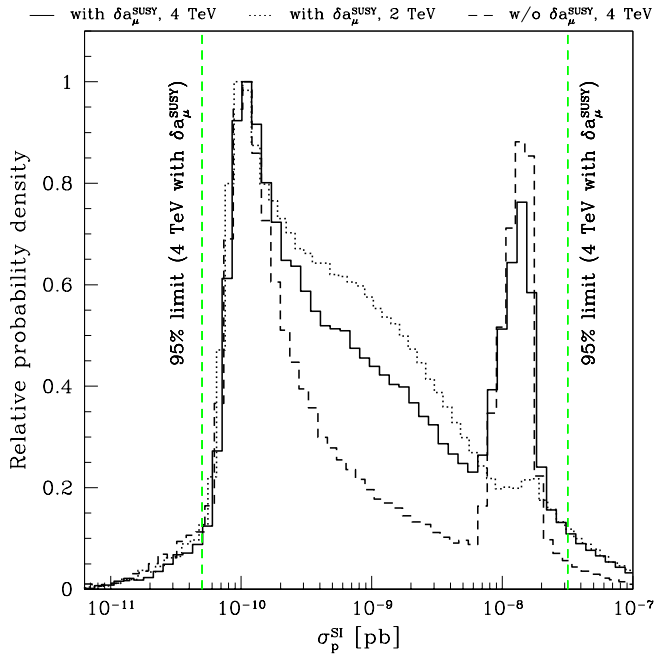}
\caption{The 1--dimensional probability distribution for the
spin--independent neutralino scattering cross section in the
CMSSM. The three curves explore the impact of different
assumptions in the analysis: assuming a $4 \TeV$ upper prior on
$m_0$ (solid), a $2 \TeV$ prior (dotted) or removing from the
analysis the constrain from $g-2$ (dashed, for the $4 \TeV$ prior
case) The vertical, dashed lines delimit the region encompassing
95\% probability for the $4 \TeV$ prior and including $g-2$. The
spike around $\sigsip \sim 10^{-8}$ corresponds to the focus point
region at large $m_0$. Future one--tonne detectors will reach down
to $\sigsip \sim 10^{-10}$ and thus explore most of the $95\%$
region.} \label{fig:1d}
\end{figure}

After marginalizing over all other parameters, we obtain the
following 1--dimensional regions encompassing 68\% and 95\% of the
total probability (see Fig.~\ref{fig:1d}):
\be \label{eq:hpr}
 \begin{aligned}
  1.0\times10^{-10}\pb   < \sigsip < 1.0\times10^{-8}\pb \quad & (68\% \text{ region}), \\
  0.5\times10^{-10}\pb < \sigsip < 3.2\times10^{-8}\pb \quad & (95\% \text{ region}).
 \end{aligned}
\ee

Currently running experiments (most notably CDMS--II but also
Edelweiss--II and ZEPLIN--II) should be able to reach down to a
few $\times 10^{-8}\pb$, on the edge of exploring this FP region.
A future generation of ``one--tonne'' detectors is going to reach
down to $\sigsip\gsim 10^{-10}\pb$, thus exploring almost the
whole $68\%$ region and much of the $95\%$ interval as well.

This result for the pdf on $\sigsip$ is rather robust with respect
to a change in the prior range. The spike at $\sigsip \sim
10^{-8}$ disappears if one cuts away most of the focus point by
imposing a prior $m_0 \leq 2 \TeV$, but the $95\%$ region is only
slightly shifted to lower values. We have also considered the
impact of removing  from the analysis the constrain coming from
$g-2$. We have found that in this case $m_0$ becomes essentially
unconstrained, and this adds statistical weight to the focus point
region. However, even in this case the high--probability range for
$\sigsip$ remains close to the values given in Eq.~\eqref{eq:hpr},
shifting to $0.3\times10^{-10}\pb < \sigsip < 2.8\times10^{-8}\pb$
($95\%$ probability region).

We notice that the high--probability regions as described above do
not necessarily coincide with the best fitting points in parameter
space if the pdf is strongly non--Gaussian, as in the present
case. We refer the reader to~\cite{rrt1} for a detailed
description of the discrepancy and a discussion of its meaning in
terms of probabilistic inference.

\begin{figure}[tb]
\centering
\includegraphics[width=\linewidth]{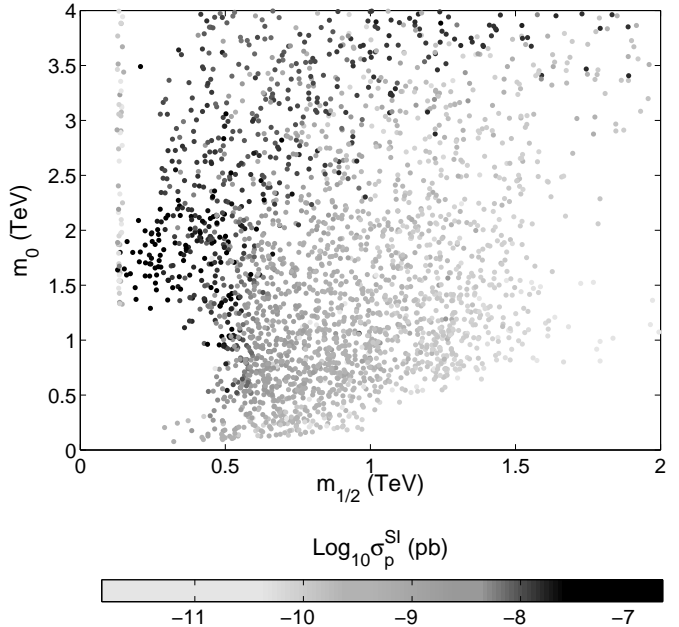}
\caption{Distribution of the values of $\sigsip$ in the $(\mhalf,
\mzero$) plane. The dots delimit the 95\% confidence region in
parameter space. Values down to $\sigsip \sim 10^{-8}$ (darker
dots) will be among the first to be explored by currently running
dark matter searches. This includes a large parte of the focus
point region at large $m_0$.} \label{fig:3d}
\end{figure}
Finally, it is interesting to consider the complementarity between
direct dark matter searches and the reach of future collider
experiments, especially in view of the imminent start of
operations at LHC. We plot in Fig.~\ref{fig:3d} the distribution
of values of $\sigsip$ in the plane $(\mhalf, \mzero)$, where the
points roughly cover the region encompassing $95\%$ of
probability. It is apparent that the largest values of the cross
section ($\sigsip \gsim 10^{-8}$)  are clustered in the region
$200 \GeV \lsim \mhalf \lsim 500 \GeV$, covering values of
$\mzero$ all the way up to our prior limit of $4 \TeV$ in the
focus point region. As the currently running dark matter detectors
start probing value down to $\sigsip \sim 10^{-8}$, they will
begin exploring the focus point region of parameter space, which
lies at large values of $\mzero$ that are largely outside the
reach of LHC searches. From our analysis a very promising synergy
thus emerges between direct dark matter collider searches, that
will soon begin to squeeze from different directions the current
high--probability region of the CMSSM parameters space.

\section{Conclusions}

We have presented a detailed investigation of the prospects for
dark matter detection in the framework of the CMSSM parameter
using state--of--the--art Bayesian methods. The power and
flexibility of the approach allows to probe many previously
unexplored ranges of parameters and to fully incorporate the
effects of remaining uncertainties in relevant SM parameters and
other theoretical uncertainties in computing mass spectra and
observables.

We have shown that the WIMP dark matter direct detection elastic
scattering cross section $\sigsip$ presents a wide spread of
values (below today's limits) at around $10^{-9\pm1}\pb$ and a
strong anticorrelation with $\mchi$. In addition, a region at
relatively large $\sigsip\simeq 1.6\times10^{-8}\pb$, and fairly
independent of $\mchi$, appears to be a feature of the focus point
region (despite large theoretical uncertainties) and will be the
first to be tested in direct detection experiments.

Despite the remaining uncertainties arising from the local dark
matter density profile and the large errors on nuclear form
factors, it seems that a large fraction of the high--probability
parameter space of neutralino dark matter in the CMSSM will be
within reach of currently running upgraded dark matter detector,
and most of it will be explored by future one--tonne detectors. We
have highlighted the complementarity of these searches with direct
SUSY searches at colliders and in particular showed that the focus
point region, mostly outside the reach of the LHC, will be among
the first to be probed by direct dark matter detection
experiments. Those results are largely robust with respect to
changes in the {\em a priori} allowed range for the parameters or
to the exclusion of the anomalous magnetic moment of the muon
measurement from the analysis.

{\em Acknowledgements:} R.T.\ is supported by the Royal
Astronomical Society through the Sir Norman Lockyer Fellowship.
R.Rda is supported by the program ``Juan de la Cierva'' of the
Ministerio de Educaci\'{o}n y Ciencia of Spain. R.Rda and R.T.
would like to thank the European Network of Theoretical
Astroparticle Physics ILIAS/N6 under contract number
RII3-CT-2004-506222 for financial support.

\end{document}